\documentclass[twocolumn,showpacs,preprintnumbers,prl]{revtex4-1}

\usepackage{graphicx}
\usepackage{amssymb}
\usepackage{epstopdf}
\usepackage{bm}

\begin{document}
    
\title{A Precision Measurement of the Neutron Scattering Length of $^4$He Using Neutron Interferometry}
    
\author{R.~Haun}
\affiliation{Department of Physics and Engineering Physics, Tulane University, New Orleans, LA 70118}
\author{M.~Arif}\thanks{deceased}
\affiliation{National Institute of Standards and Technology, Gaithersburg, MD 20899, USA}
\author{T.~C.~Black}
\affiliation{Department of Physics, University of North Carolina-Wilmington, Wilmington, NC 28403, USA}
\author{B.~Heacock}
\affiliation{Department of Physics, North Carolina State University, Raleigh, NC 27695, USA}
\affiliation{Triangle Universities Nuclear Laboratory, Durham, North Carolina 27708, USA}
\author{M.~G.~Huber}
\affiliation{National Institute of Standards and Technology, Gaithersburg, MD 20899, USA}
\author{D.~A.~Pushin}
\affiliation{Department of Physics, University of Waterloo, Waterloo, ON, Canada N2L 3G1}
\affiliation{Institute for Quantum Computing, University of Waterloo, Waterloo, ON N2L 3G1, Canada}
\author{C.~B.~Shahi}
\affiliation{Department of Physics, University of Maryland, College Park, MD 20742, USA}
\author{F.~E.~Wietfeldt}
\affiliation{Department of Physics and Engineering Physics, Tulane University, New Orleans, LA 70118}

\date{\today}

\begin{abstract}
We report a 0.08 \% measurement of the bound neutron scattering length of $^4$He using neutron interferometry. The result is 
$b = (3.0982 \pm 0.0021 \mbox{ [stat]} \pm 0.0014 \mbox{ [sys]}) \mbox{ fm}$. The corresponding free atomic scattering length is $a = (2.4746 \pm 0.0017 \mbox{ [stat]} \pm 0.0011 \mbox{ [sys]}) \mbox{ fm}$.
With this result the world average becomes $b = (3.0993 \pm 0.0025)$ fm, a 2 \% downward shift and a reduction in uncertainty by more than a factor of six. Our result is in disagreement with a previous neutron interferometric measurement but is in good agreement with earlier measurements using neutron transmission.
\end{abstract}

\maketitle
In the zero-energy limit the neutron-nucleus interaction potential can be treated as a delta function multiplied by a constant with dimension length, the neutron free scattering length $a$, which is in general spin-dependent and complex. In the case of a solid target the atom is constrained from recoiling so the bound scattering length $b = a (A + 1) / A$, where $A$ is the atom/neutron mass ratio, is used.
The neutron scattering length of an isotope determines its low energy neutron scattering and absorption cross sections.  Neutron scattering lengths are fundamental in neutron scattering applications and are widely used in neutron science and nuclear engineering. They provide a benchmark for few-body nucleon potential models and chiral effective field theories. Precise neutron scattering lengths of noble gases are needed for short-range interaction searches using cold and ultracold neutrons \cite{Ser11,Had18}.
\par
Realistic nucleon-nucleon (NN) potentials such as the Nijmegan, CD Bonn, and AV18, when used in conjunction with exact few-body computational methods, successfully predict few-nucleon scattering amplitudes in many channels but fail to reproduce three and four body binding energies \cite{Kiev10}. These models do not accurately predict the vector analyzing power $A_y$ in a number of few-nucleon systems, including $n + d$, $p + d$, $n + ^3$He, and $p + ^3$He \cite{Cleg09}.   It has long been clear that a correct description of few-nucleon systems would require not only an NN but also a 3N force \cite{Gloc96}.  A number of 3N potential models have been created including the Tucson-Melbourne \cite{Mcke68,Coon75},  Brazilian \cite{Coel83}, and  Urbana-Illinois \cite{Pudl97, Piep01}.  These can be adjusted to match the triton and $^3$He binding energies, but they do not resolve the discrepancies between theory and experiment in the scattering data and have trouble reproducing the binding energy of $^4$He. It is tempting to think that this could be resolved by adding a 4N potential, but this would require the introduction of {\em ad hoc} repulsive terms into the potential model \cite{Kiev10}.
\par
More recently, we have seen the maturation of perturbative chiral effective field theories ($\chi$EFT), which use the symmetries of QCD in a perturbative expansion of particle momenta divided by the chiral symmetry breaking scale, $\frac{Q}{\Lambda_\chi}$, where $\Lambda_\chi$ is a mass scale appropriate to the system.  To implement $\chi$EFT to solve for nuclear forces, the long and intermediate range interactions are calculated explicitly.  The short range behavior is accounted for through use of low energy constants (LEC) that are adjusted to match experimental data \cite{Mach11}. With this prescription one can construct an NN potential that contains all terms consistent with the symmetries of the strong interaction.  The power counting scheme (power $\nu$ of $\frac{Q}{\Lambda_\chi}$ terms included) determines which exchange diagrams to include at any particular order of the calculation so  the precision of the expansion is controllable.  Leading order (LO, $\nu$ = 0) and ``next to leading order'' (NLO, $\nu$ = 1) diagrams produce two nucleon forces. Three nucleon force diagrams first appear at NNLO ($\nu$ = 2), and four nucleon diagrams at N$^3$LO  ($\nu$ = 3). These arise naturally in EFT.  Nucleon potentials have now been constructed to N$^3$LO \cite{Skib11}.  Calculations at this order are as yet unable to resolve the outstanding discrepancies between theory and experiment in few nucleon systems. It is believed that EFT potentials must be constructed at higher orders to bring them into alignment.  
\par
A significant motivation for more precise measurements of neutron scattering lengths of light nuclei, and in particular this measurement, is to provide high-quality ``set-point'' data for effective range expansions of $n$+nucleus systems.  These expansions can be used to assist construction of improved realistic 3N and 4N potential models, and help constrain low energy constants used in building models at higher orders in chiral effective field theory.  It is hoped that such new models will bring few nucleon theory and experiment into better agreement.  We note that a high precision neutron interferometry measurement of the $n$-$d$ scattering length \cite{Sch03} has already been used to help fix the LEC's for the N$^3$LO 3N force interaction \cite{Kiev10}. 
\par
\begin{figure}
\begin{center}
\includegraphics[width = 3.5in]{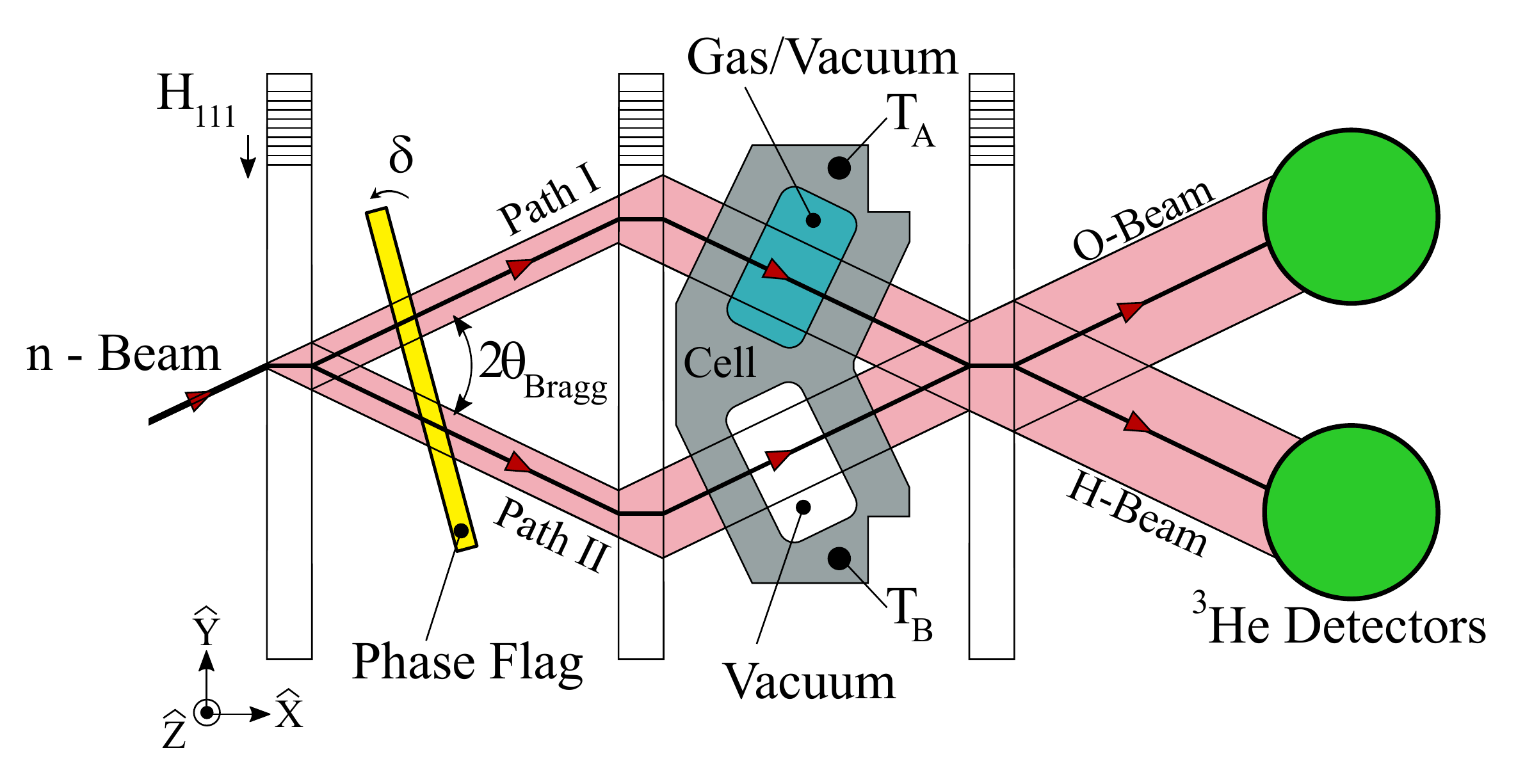}
\end{center}
\caption{\label{F:n4HeSetup} Overhead view of the experimental set up. The incident neutron wave is split into two paths by Bragg diffraction in the first blade, reflected
by the second blade, and mixed coherently in the third blade, producing interference fringes in the $^3$He detector count rates, modulated by the angle $\delta$ of the
phase flag. The target contains pressurized gas (or vacuum) in path I and vacuum in path II.}
\end{figure}

A neutron interferometer  \cite{Sea89,Kai99,Rau15} splits the matter wave of a neutron into two coherent paths using Bragg diffraction in single crystal silicon and then reflects and recombines them, producing interference that is observable by neutron counters located behind the crystal. A target placed in one beam path of the neutron interferometer produces a relative phase shift $\phi = -N \lambda \overline{b} l$ where $N$ is the atomic number density, $l$ is the target path length, $\lambda$ is the neutron wavelength, and $\overline{b}$ is the real part of the average bound neutron scattering length in the target. In a neutron interferometer the observed phase shift is due to coherent forward scattering with zero momentum transfer, so the bound scattering length is used regardless of the state of the target. This work followed a method similar to that used in previous $b$ measurements for light gases (H$_2$, D$_2$, and $^3$He) at the NIST Neutron Interferometry and Optics Facility (NIOF) \cite{Sch03,Huf04,NIOF}. The experimental set up is depicted in figure \ref{F:n4HeSetup}. A focused monochromatic neutron beam was incident on the first blade of the three blade single crystal silicon interferometer. A 1.5 mm thick fused silica phase flag mounted to a precision rotation stage intercepted both beam paths prior to the second blade. The two neutron paths then passed through the target and converged on the third blade where they interfered. Neutrons were detected and counted by two $^3$He proportional counters with a relative probability that depended on the difference in neutron phase shifts of the two paths.

\par
The target was a double cell constructed of 6061 alloy aluminum. When in place for a gas phase shift measurement, one neutron beam path passed through the gas-filled cell and the other through the evacuated cell. The target geometry was designed so that each path passed perpendicularly through the cell interfaces.  The purpose of the evacuated cell was to equalize the neutron phase shift in aluminum between the beam paths which maximizes the fringe contrast by preventing decoherence. Any small relative difference in the aluminum phase shift was accounted for by the empty cell measurement described below. The aluminum target lid (not shown) was sealed to the cell body with an indium gasket. The target was attached to a kinematic mount that was suspended from above by a system of precision computer-controlled translation and rotation stages. To align the target, a fused silica alignment slab was precisely aligned to it mechanically and inserted into the interferometer on the kinematic mount. The alignment slab was rotated about the vertical and transverse axes to equalize the neutron path length through it, as measured by the minimum in the phase shift difference of the two paths. This reduced the phase shift error in the target due to angular misalignment to $< 1$ mrad.
\par
The target gas was supplied by Matheson TriGas \cite{DISCLM,Mat} and had a certified atomic purity of 99.9999 \% natural helium. In addition to $^4$He, the main components were $^3$He ($2\times 10^{-4}$ \% natural abundance) and $^{14}$N (about $1\times 10^{-4}$ \%). A stainless steel gas handling system, with VCR \cite{VCR} valves and fittings and a turbomolecular vacuum pump, was used to evacuate and fill the target. Gas pressure in the cell was measured by a Paroscientific Digiquartz 745 precision sensor \cite{Paro}, calibrated at NIST to an uncertainty of $\pm 22$ Pa ($\pm0.22$ mbar)\cite{sig}. Vacuum was measured using an ionization gauge. Prior to the experiment, the gas handling system and both target cells were evacuated to a pressure of $1.1 \times 10^{-3}$ Pa ($8 \times 10^{-6}$ torr), flushed with pure nitrogen and helium several times each, and evacuated again. A residual gas analyzer measured 60 \% H$_2$O, 16 \% H$_2$, 12 \% N$_2$, and 10 \% CO$_2$. Cell temperature during the experiment was measured using a pair of precision thermistors, NIST calibrated to $\pm 1.2$ mK, imbedded in the target lid.
\par
\begin{figure}
\begin{center}
\includegraphics[width = 3.5in]{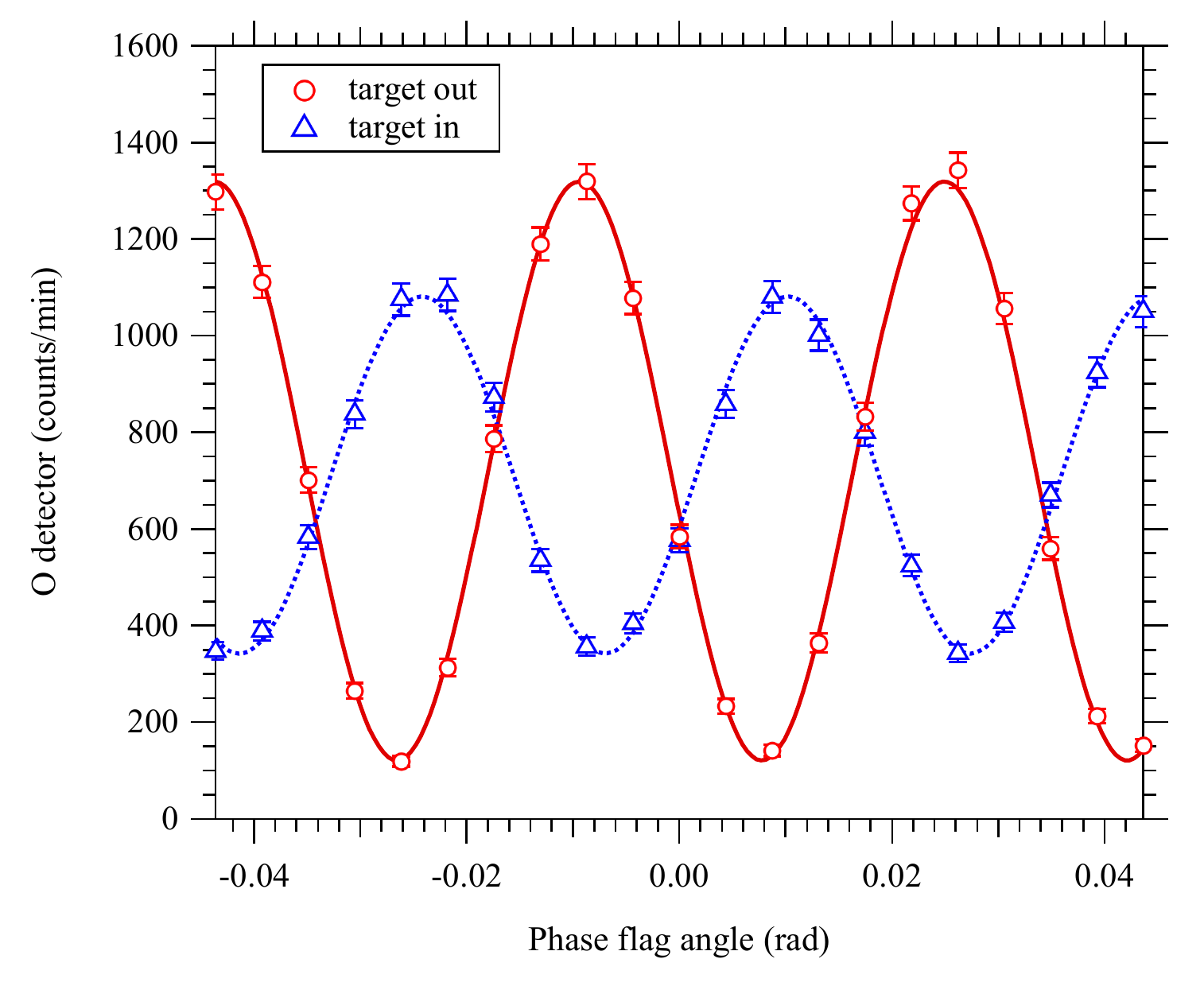}
\end{center}
\caption{\label{F:interf} Typical interferograms from the 12.9 bar data set, {\em i.e.} counts in the O-detector {\em vs.} phase flag angle, for target out and in. Error bars are statistical.}
\end{figure}

The measurement procedure was as follows. The gas cell was filled to the desired pressure and the target was inserted into the interferometer. The phase flag was rotated in 20 steps over a range of $\pm$ 2.5 degrees. This produced an interferogram, a cosine interference function of the neutron count rate in the O-beam neutron counter caused by the difference in relative neutron path length through the phase flag as it rotated. Each interferogram required 21 minutes to complete. The target was then translated out of the interferometer and another interferogram was taken to measure the intrinsic phase $\phi_0$ associated with the interferometer setup absent the target. These measurements were repeated in opposite order to produce a four step sequence: target out, in, in, out. The net phase difference, target in ($\phi_{\rm full}$) minus target out ($\phi_0$), gives the neutron phase shift due to the target $\Phi_D  = \phi_{\rm full} - \phi_0$ while canceling any first order drift in $\phi_0$. Typical target in and target out interferograms are shown in figure \ref{F:interf}.
\par
\begin{figure}
\begin{center}
\includegraphics[width = 3.5in]{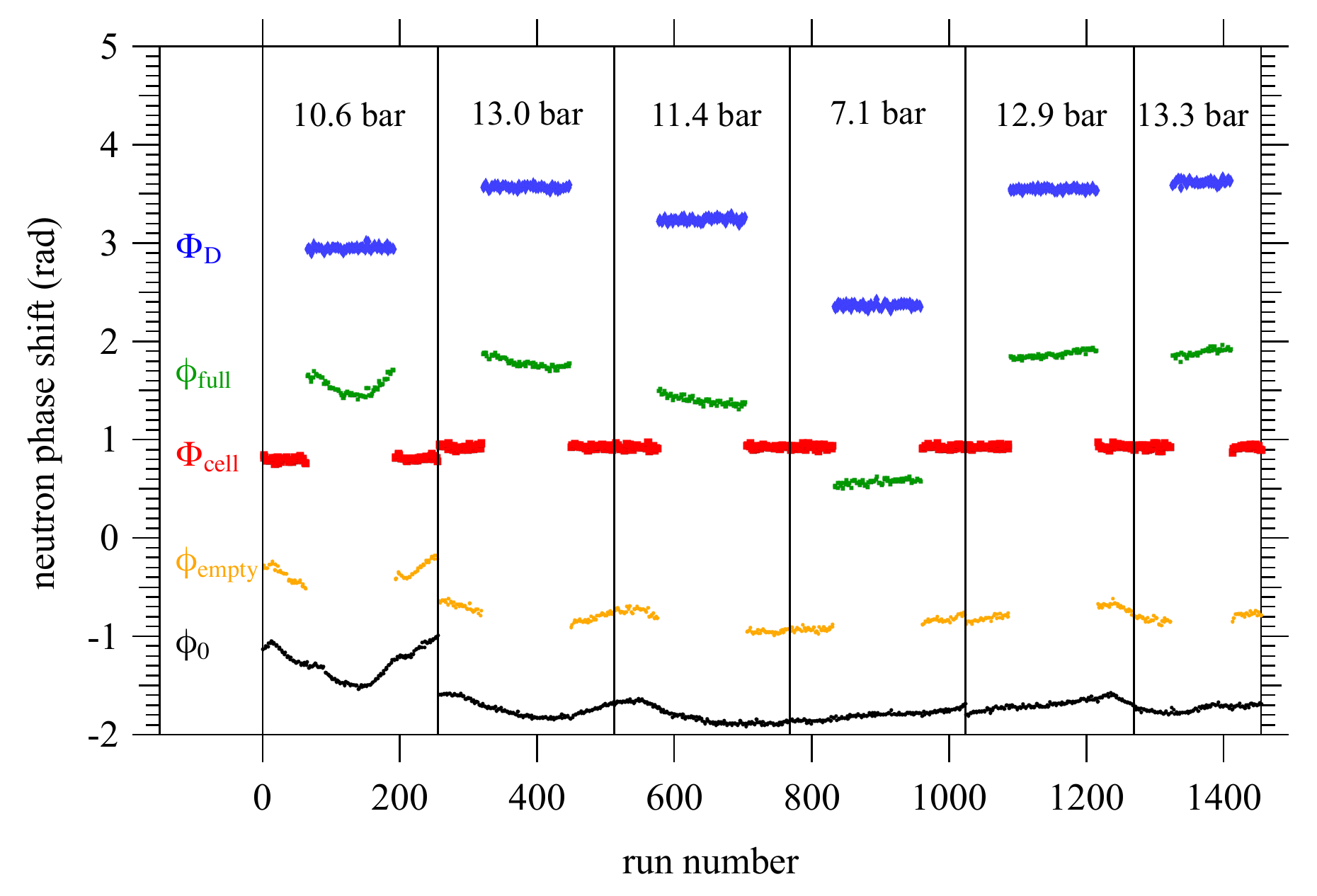}
\end{center}
\caption{\label{F:allPhase} The total data set,  neutron phase from 1456 interferograms for target-in with gas ($\phi_{\rm full}$, green), target-in empty ($\phi_{\rm empty}$, gold) and target-out ($\phi_0$, black). The blue points are $\Phi_D = \phi_{\rm full} -\phi_0$. The red points are $\Phi_{\rm cell} = \phi_{\rm empty} -\phi_0$. Subtracting $\Phi_{\rm cell}$ from $\Phi_D$ gives the phase shift $\Phi_{\rm gas}$ due to the gas only. Vertical lines delineate the runs used to compute $\Phi_{\rm gas}$ for each pressure data set. The discontinuous jumps in $\phi_0$ and $\phi_{\rm empty}$ between 10.6 bar and 13.0 bar were due to a gap in time of about one month and an improvement in the facility's environmental controls. All phase shifts are shown modulo $2\pi$.}
\end{figure}

Data sets were collected using six different helium pressures in order to investigate any pressure-dependent systematic effects. Before and after each helium data set, a set of empty target phase measurements ($\phi_{\rm empty}$, both target cells evacuated) was taken, using the same four step sequence described above, to measure the neutron phase shift $\Phi_{\rm cell} = \phi_{\rm empty} - \phi_0$ due to the empty aluminum target. The neutron phase shift attributed to the helium was then $\Phi_{\rm gas} = \Phi_D - \Phi_{\rm cell}$. A systematic problem can arise from the fact that the target's temperature may differ from the interferometer crystal temperature, and it tends to rise slowly over the course of a data set as it is translated in and out by the motor-driven stage. The interferometer is very sensitive to thermal gradients so this changes the intrinsic phase of the interferometer. As a result the actual $\Phi_{\rm cell}$ may drift in time and differ significantly between the gas-filled and empty-cell phase measurements. This problem confounded an earlier attempt at this experiment. Our solution was to attach a glycol-cooled copper block to the target's translation motor. By varying the glycol temperature we found an operating value that reduced the time-variation in $\Phi_{\rm cell}$ to a negligible level. This was verified using a ``dummy'' target; an aluminum target of similar construction with through holes for the neutron beam paths to remove the neutron phase shift in the target and isolate the temperature gradient effect. Figure \ref{F:allPhase} shows the fitted phase of all 1456 interferograms taken during the experiment, separating target-in, target-out, and empty target-in measurements. For each gas pressure, an equal number of interferograms was collected with the target full ($\phi_{\rm full}$) and the target empty ($\phi_{\rm empty}$), with the latter divided equally between before and after, to cancel any small linear drift in $\Phi_{\rm cell}$. 
\par
The $^4$He bound scattering length $b$ was then calculated using
\begin{equation}
\label{E:bMeas}
b = \frac{\Phi_{\rm gas}}{N \lambda D}.
\end{equation}
The $^4$He density $N$ was calculated from the measured pressure ($P$) and temperature ($T$) using the virial equation
\begin{equation}
N(T,P) = \frac{P}{k_B T \left( 1 + B_P + C_P P^2 \right)}
\end{equation}
where $k_B$ is the Boltzmann constant and $B_P$, $C_P$ are the tabulated virial coefficients for helium \cite{Dym02}. The neutron wavelength 
$\lambda$ = (2.70913 $\pm$ 0.00016 [stat] $\pm$ 0.00023 [sys]) was measured in the O-beam at the exit of the neutron interferometer using a standard
Bragg diffraction rocking curve method (see for example \cite{Lit97}) with a pressed silicon crystal. The neutron path length $D$ through the gas target was measured 
at the NIST Precision Engineering Division Coordinate Measuring Machine \cite{CMM} to be $D = (1.0016 \pm 0.0001)$ cm, unpressurized at 20 $^\circ$C. The entrance and exit windows were nominally 0.6 cm thick and measured to 180 nm precision. The change in thickness of gas and aluminum due to target deformation when pressurized was calculated using finite element analysis in Autodesk Inventor \cite{Auto}. The dominant systematic effect is the difference in relative path lengths in aluminum. We found that the change in relative path length (pressurized cell {\em vs.} evacuated cell) was 190 nm (0.003 \%) at 13 bar. The corresponding proportional correction was applied to $\Phi_{\rm gas}$ for each pressure.
\begin{figure}
\begin{center}
\includegraphics[width = 3.5in]{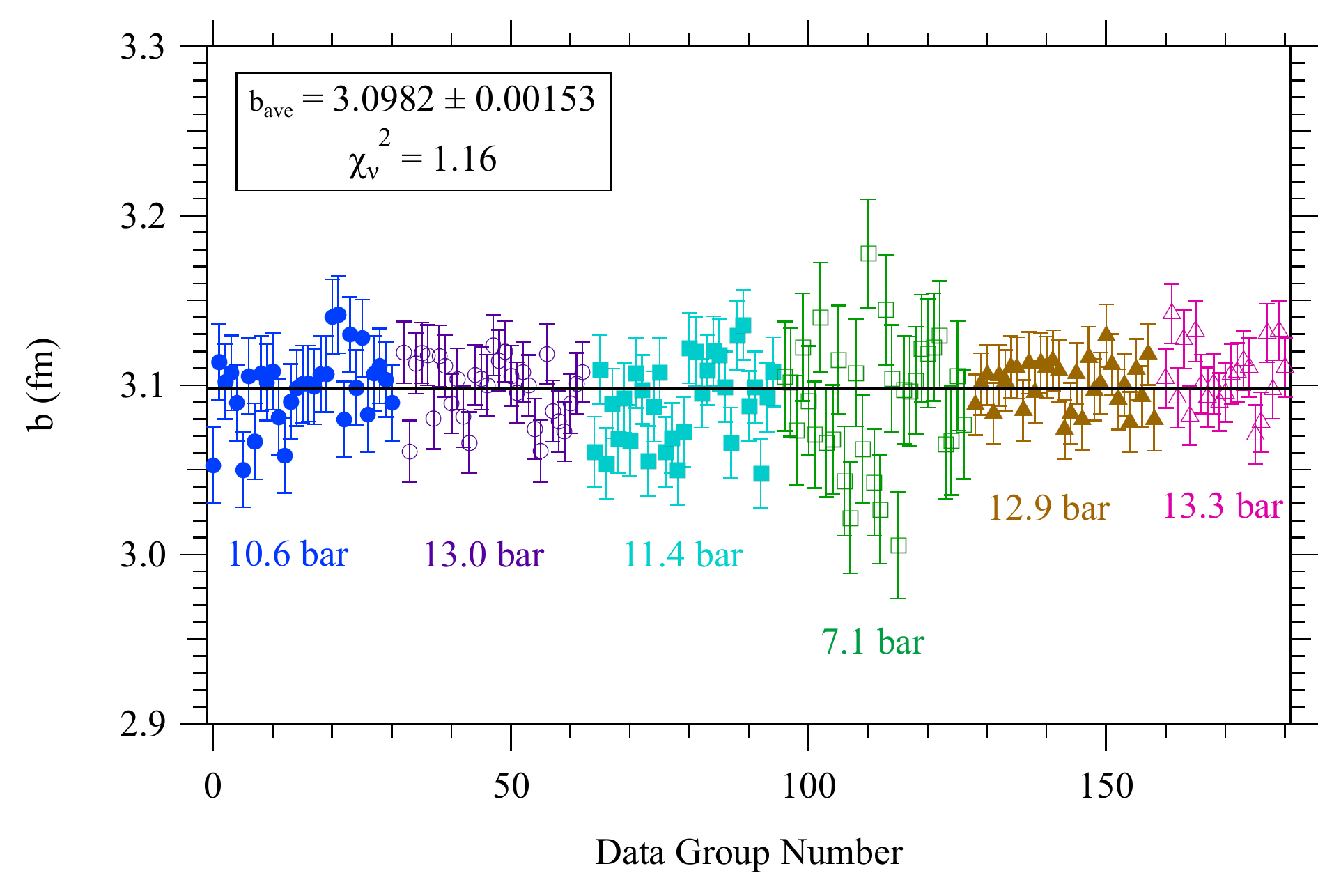}
\end{center}
\caption{\label{F:bFit} Bound neutron scattering length for the six data sets after all corrections, fit to a constant. Error bars are due to Poisson counting statistics from the
$\Phi_D$ data only.}
\end{figure}
\par
\begin{figure}
\begin{center}
\includegraphics[width = 3.5in]{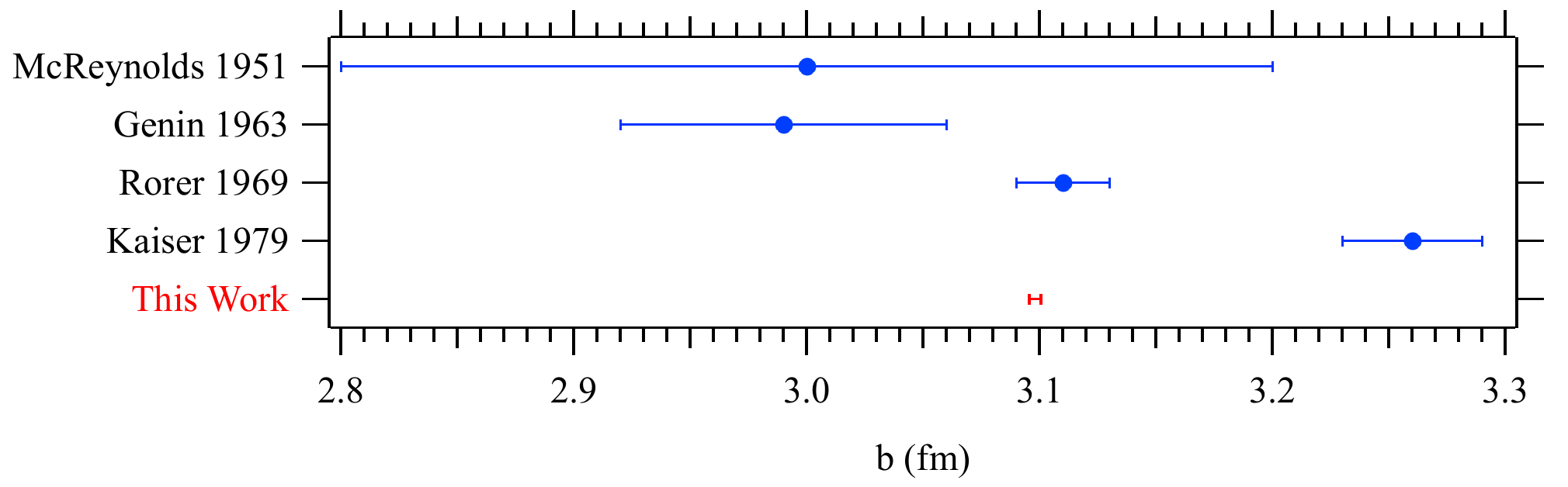}
\end{center}
\caption{\label{F:sumPlot} A summary showing this result in comparison to previous transmission measurements by McReynolds \cite{McR51}, Genin {\em et al.} \cite{Gen63},
Rorer {\em et al.} \cite{Ror69}, and neutron interferometry by Kaiser {\em et al.} \cite{Kai79}}
\end{figure}

The values of $b$ found by applying equation \ref{E:bMeas} to each $\Phi_{\rm gas}$ measurement for the six data sets, with all corrections applied, are shown in figure \ref{F:bFit}, along with the weighted average. Our final result for the bound $n-^4$He scattering length is
\begin{equation}
b = (3.0982 \pm 0.0021 \mbox{ [stat]} \pm 0.0014 \mbox{ [sys]}) \mbox{ fm}
\end{equation}
or expressed as the free scattering length
\begin{equation}
a = (2.4746 \pm 0.0017 \mbox{ [stat]} \pm 0.0011 \mbox{ [sys]}) \mbox{ fm}.
\end{equation}
The error budget is shown in table \ref{T:Errors}. The largest systematic correction and uncertainty was due to the target cell deformation calculation.  The total statistical uncertainty includes contributions from the gas-filled phase shifts (see figure 4) and the empty cell measurements. Our result is in disagreement with the previous neutron interferometric measurement of Kaiser {\em et al.} \cite{Kai79}, but in good agreement with earlier measurements that used the transmission method \cite{McR51,Gen63,Ror69} (see figure \ref{F:sumPlot}). Including this measurement, the world average becomes
$b = (3.0993 \pm 0.0025)$ fm, a 2 \% downward shift and a reduction in the net uncertainty by a factor of more than six.
\par
We gratefully acknowledge support from the National Science Foundation grant PHY-1205342 and NIST (US Department of Commerce), and the NCNR for providing the neutron facilities used in this work and for technical support. We thank Christoph Brocker for his assistance in FEA modeling and Tobias Herman for his work on sensor calibrations.
\begin{table}
\caption{\label{T:Errors} A summary of systematic corrections and uncertainties. The statistical uncertainty is the combined Poisson counting statistics from $\Phi_D$ (see figure \ref{F:bFit}) and $\Phi_{\rm cell}$ data.}
\centering
\begin{ruledtabular}
\begin{tabular}{lll}
& correction (fm) & 1 $\sigma$ uncertainty (fm)\\\hline
target cell deformation & 0.01249 & 0.00129\\
target cell metrology & & 0.00031\\
neutron wavelength &  & 0.00032\\
gas pressure cal. &  & 0.00032\\
gas temperature cal. & & 0.00001\\
virial coefficients & & 0.00021\\
gas purity & -0.00002 & 0.00001\\\hline
total systematic & 0.01247 & 0.00142\\
statistical & & 0.00214\\\hline
\end{tabular}
\end{ruledtabular}
\end{table}


\begin{thebibliography}{99}
\bibitem{Ser11} A.~P.~Serebrov, {\em et al.}, Phys.~Rev.~C {\bf 84}, 044001 (2011).
\bibitem{Had18} C.~C.~Haddock, {\em et al.}, Phys.~Rev.~D {\bf 97}, 062002 (2018).
\bibitem{Kiev10}  A.~Kievsky, M.~Viviani, L.~Girlanda, and L.~E.~Marcucci, Phys. Rev. C 81, 044003 (2010).
\bibitem{Cleg09} T.~Clegg, Proc. of Sci. CD09:61 (2009).
\bibitem{Gloc96} W.~Gl\"{o}ckle, H.~Witala, D.~H\"{u}ber, H.~Kamada, and J.~Golak,  Phys. Rept. {\bf 274}, 107 (1996).  
\bibitem{Mcke68}  B.~H.~J.~McKellar and R.~Rajaraman, Phys. Rev. Lett. {\bf 21}, 450 (1968).
\bibitem{Coon75} S.~A.~Coon, M.~D.~Scadron and B.~R.~Barrett, Nucl. Phys. {\bf A242}, 467 (1975).
\bibitem{Coel83} H.~T.~Coelho, T.~K.~Das and M.~R.~Robilotta, Phys. Rev. C {\bf 28}, 1812 (1983).
\bibitem{Pudl97} B.~S.~Pudliner, V.~R.~Pandharipande, J.~Carlson, S.~C.~Pieper and R.~B.~Wiringa, Phys. Rev. C {\bf 56}, 1720 (1997).
\bibitem{Piep01} S.~C.~Pieper, V.~R.~Pandharipande, R.~B.~Wiringa and J.~Carlson, Phys. Rev. C {\bf 64}, 014001 (2001).
\bibitem{Mach11} R.~Machleidt and D.~R.~Entem, Physics Reports 503 (2011).  
\bibitem{Skib11} R. Skibi\'{n}ski, J. Golak, K. Topolnicki, H. Wita\l{}a, E. Epelbaum, W. Gl\"{o}ckle, H. Krebs, A. Nogga, and H. Kamada, Phys. Rev. C 84, 054005 (2011).  
\bibitem{Sch03} K.~Schoen, {\em et al.}, Phys. Rev. C {\bf 67}, 044005 (2003).
\bibitem{Sea89} V.~F.~Sears, {\em Neutron Optics}, Oxford University Press (1989).
\bibitem{Kai99} H.~Kaiser and H.~Rauch, {\em De Broglie wave optics: neutrons, atoms and molecules}, in Bergmann/Schaefer Optics of Waves and Particles, Walter de Gruyter Verlag (1999).
\bibitem{Rau15} H.~Rauch and S.~A.~Werner, {\em Neutron Interferometry}, 2nd ed., Oxford Science Publications (2015).
\bibitem{Huf04} P.~Huffman, {\em et al.}, Phys. Rev. C {\bf 70}, 014004 (2004).
\bibitem{NIOF} {\tt https://www.nist.gov/laboratories/tools-instruments/}\\{\tt neutron-interferometry-and-optics-facility-niof}
\bibitem{Mat} {\tt https://www.mathesongas.com/gases}
\bibitem{DISCLM} Certain trade names and company products are 
mentioned in the text or identified in illustrations in order to 
adequately specify the experimental procedure and equipment used. In 
no case does such identification imply recommendation or endorsement 
by the National Institute of Standards and Technology, nor does it imply 
that the products are necessarily the best available for the purpose.
\bibitem{VCR} {\tt https://www.swagelok.com/en/product/Fittings/}\\{\tt VCR-Metal-Gasket-Face-Seal}
\bibitem{Paro} \verb=http://paroscientific.com/pdf/D75_Model_745_765.pdf=
\bibitem{sig} Throughout this letter all stated uncertainties are at the 68.3 \% confidence level.
\bibitem{Dym02} J. H. Dymond, K. N. Marsh, R. C. Wilhoit, M. Wong, K. C. Edited by Frenkel, and K. N. Marsh. {\em Virial Coefficients of Pure Gases and Mixtures}, Landolt-B\"{o}rnstein, New Series IV/21A, Springer (2002).
\bibitem{Lit97} K.~C.~Littrell, B.~E.~Allman, and S.~A.~Werner, Phys.~Rev.~A {\bf 56}, 1767 (1997).
\bibitem{CMM} J. Stoup. NIST test No. 821/265253-01.
\bibitem{Auto} {\tt https://www.autodesk.com/products/inventor/features}
\bibitem{Kai79} H.~Kaiser, {\em et al.}, Zeit. Phys. A, {\bf 291}, 231 (1979).
\bibitem{McR51} A.~W.~McReynolds, Phys. Rev. {\bf 84}, 969 (1051).
\bibitem{Gen63} R.~Genin, {\em et al.}, J. Phys. Radium, {\bf 24}, 21 (1963).
\bibitem{Ror69} D.~C.~Rorer, B.~M.~Ecker, and R.~\"{O}.~Aky\"{u}z, Nucl. Phys. {\bf A133}, 410 (1969).

\end{thebibliography}
\end{document}